\documentclass[aps,pra,reprint,showpacs,superscriptaddress,twocolumn]{revtex4-1}

\usepackage{amsmath}
\usepackage{amssymb}
\usepackage{graphicx}
\usepackage{hyperref}

\hypersetup{
    colorlinks=true,linkcolor=blue,citecolor=blue,
    filecolor=blue,urlcolor=blue,breaklinks=true
}

\RequirePackage{color}

\begin{document}

\title{Scattering and bound states of a spin--1/2 neutral particle in
the cosmic string spacetime}

\author{Fabiano M. Andrade}
\email{fmandrade@uepg.br}
\affiliation{
 Department of Computer Science
 and Department of Physics and Astronomy,
 University College London,
 WC1E 6BT London, United Kingdom
}
\affiliation{
  Departamento de Matem\'{a}tica e Estat\'{i}stica,
  Universidade Estadual de Ponta Grossa,
  84030-900 Ponta Grossa-PR, Brazil
}

\author{Cleverson Filgueiras}
\email{cleversonfilgueiras@yahoo.com.br}
\affiliation{
  Departamento de F\'{i}sica,
  Universidade Federal de Lavras, Caixa Postal 3037,
  37200-000, Lavras-MG, Brazil
}

\author{Edilberto O. Silva}
\email{edilbertoo@gmail.com}
\affiliation{
  Departamento de F\'{i}sica,
  Universidade Federal do Maranh\~{a}o,
  65085-580 S\~{a}o Lu\'{i}s-MA, Brazil
}

\date{\today}

\begin{abstract}
In this paper the relativistic quantum dynamics of a spin-1/2 neutral
particle with a magnetic moment $\mu$ in the cosmic string spacetime is
reexamined by applying the  von Neumann theory of self--adjoint
extensions.
Contrary to previous studies where the interaction between the spin and
the line of charge is neglected, here we consider its effects.
This interaction gives rise to a point interaction:
$\boldsymbol{\nabla} \cdot \mathbf{E}= (2\lambda/\alpha)\delta(r)/r$.
Due to the presence of the Dirac delta function, by applying
an appropriated boundary condition provided by the theory of
self--adjoint extensions, irregular solutions for the Hamiltonian are
allowed.
We address the scattering problem obtaining the phase shift, S-matrix
and the scattering amplitude.
The scattering amplitude obtained shows a dependency with energy which
stems from the fact that the helicity is not conserved in this
system.
Examining the poles of the S-matrix we obtain an expression for the
bound states.
The presence of bound states for this system has not been discussed
before in the literature.
\end{abstract}

\pacs{03.65.Ge, 03.65.Db, 98.80.Cq, 03.65.Pm}
\maketitle

\section{Introduction}

Theory of topological defects is a natural framework for studying
properties of physical systems.
In cosmology, the origin of defects can be understood as a sequence of
phase transitions in the early universe.
These processes occur with critical temperatures which are related to
the corresponding symmetry spontaneously breaking scales
\cite{PRD.1974.9.3320,PRD.1974.9.3357,RPP.1979.42.389}.
These phase transitions can give rise to topologically stable defects,
for example, domain walls, strings and monopoles
\cite{JPA.1976.9.13871976}.
Topological defects are also found in condensed matter systems.
In these systems, they appears as vortices in superconductors, domain
wall in magnetic materials, dislocations of crystalline substances,
among others.
An important property that can be verified in topological defects is
that they are described by a spacetime metric with a
Riemann--Christoffel curvature tensor which is null everywhere except on
the defects.
Here, we look for a cosmic string, which is a linear
topological defect with a conical singularity at the origin.
The interest in this subject has contributed to the understanding and
advancement of other physical phenomena occurring in the universe and
also in the context of non-relativistic physics.
For example, in the galaxy formation
\cite{PRL.1984.53.1700,PRD.1986.33.2175}, to study vortex
solutions in non-abelian gauge theories with spontaneous symmetry
breaking \cite{PLB.1986.171.199} and to study the gravitational analogue
of the Aharonov--Bohm effect
\cite{NC.1967.52.129,JPA.1981.14.2353,PRD.1982.26.1281,
PRD.1987.35.2031,AoP.1989.193.142}.
In recent developments, cosmic strings have been considered to analyze
solutions in de Sitter and anti-de Sitter spacetimes
\cite{PRD.2016.94.063524}, to study the thermodynamic properties of a
neutral particle in a magnetic cosmic string background by using an
approach based on the partition function method \cite{EPJC.2016.76.553},
to compute the vacuum polarization energy of string configurations in
models similar to the standard model of particle physics
\cite{PRD.2016.94.045015}, to find the deflection angle in the weak
limit approximation by a spinning cosmic string in the context of the
Einstein--Cartan theory of gravity \cite{EPJC.2016.76.332}, to analyze
numerically the behavior of the solutions corresponding to an Abelian
string in the framework of the Starobinsky model
\cite{CQG.2016.33.055004}, to study solutions of black holes
\cite{AoP.2015.362.576}, to investigate the average rate of change of
energy for a static atom immersed in a thermal bath of electromagnetic
radiation \cite{PRD.2015.92.084062}, to study Hawking radiation of
massless and massive charged particles \cite{JRG.2015.47.1}, to study
the non-Abelian Higgs model coupled with gravity
\cite{CQG.2015.32.105001}, in the quantum dynamics of scalar bosons
\cite{EPJC.2015.75.287}, hydrodynamics \cite{AJ.2015.804.121}, to study
the non-relativistic motion of a quantum particle subjected to magnetic
field \cite{AoP.2015.356.346}, to investigate dynamical solutions in the 
context of super--critical tensions \cite{PRD.2015.91.064010}, Higgs
condensate \cite{PRD.2015.91.043001}, to analyze the effects on spin
current and Hall electric field
\cite{PRA.2013.87.032107,PRD.2014.90.125014}, to investigate the
dynamics of the Dirac oscillator
\cite{PRA.2011.84.32109,EPJC.2014.74.3187}, to study non-inertial effects
on the ground state energy of a massive scalar field
\cite{PRD.2014.89.027702}, Landau quantization \cite{AoP.2014.350.105}
and to investigate the quantum vacuum interaction energy
\cite{PRD.2014.89.065034}.

In the present work, we study the quantum dynamics of a spin--1/2
neutral particle in the presence of an electric field due to an
infinitely long, infinitesimally thin line of charge along the $z$--axis
of the cosmic string, with constant charge density on it.
This model have been studied in Ref. \cite{JHEP.2004.2004.16} in the
non-relativistic regime and, for this particular case, only the
scattering problem was considered.
The present system is an adaptation of the usual Aharonov-Casher problem
\cite{PRL.1984.53.319} (which is dual to the Aharonov-Bohm problem
\cite{PR.1959.115.485}), where now effects of localized curvature are
included in the model.
We reexamine this problem by using the von Neumann theory of
self--adjoint extensions \cite{JMP.1985.26.2520,Book.2004.Albeverio}.
We address the relativistic case and investigate some
questions that were not considered in the previous studies, as for
example, the existence of bound states.
For this, we solve the scattering problem and derive the $S$ matrix in
order to obtain such bound states.

The plan of this work is the following.
In Section \ref{motion-equation}, we derive the Dirac-Pauli equation in
the cosmic string spacetime without neglecting the term which depends
explicitly on the spin.
Arguments based on the theory of self--adjoint extension are given in
order to make clear the reasons why we should consider the spin effects
in the dynamics of the system.
In Section \ref{bound-scatt}, we study the Dirac--Pauli Hamiltonian via
the von Neumann theory of self--adjoint extension.
We address the scattering scenario within the framework of Dirac--Pauli
equation.
Expressions for the phase shift, S-matrix, and bound states are
derived.
We also make an investigation on the helicity conservation problem in
the present framework.
A brief conclusion is outlined in Section \ref{conclusion}.

\section{The relativistic equation of motion}
\label{motion-equation}

The model that we address here consists of a spin--1/2 neutral particle
with mass $M$ and magnetic moment $\mu$, moving in an external
electromagnetic field $F_{\mu \nu}$ in the cosmic string spacetime,
described by the line element in cylindrical coordinates,
\begin{equation}
  ds^{2}=c^{2}dt^{2}-dr^{2}-\alpha^{2}r^{2}d\varphi^{2}-dz^{2},
  \label{eq:metric}
\end{equation}
with $-\infty<(t,z)<\infty $, $r \geq 0$, $0\leq \varphi \leq 2 \pi$ and
$\alpha$ is given in terms of the linear mass density $\tilde{m}$ of the
cosmic string by $\alpha =1-4\tilde{m}/c^{2}$.
This metric has a cone-like singularity at $r=0$
\cite{SPD.1977.22.312}.
In this system, the fermion  particle is described by a four--component
spinorial wave function $\Psi$ obeying the generalized Dirac--Pauli
equation in a non flat spacetime, which should include the spin
connection in the differential operator.
Moreover, in order to make the Dirac--Pauli equation valid in curved
spacetime, we must rewrite the standard Dirac matrices, which are
written in terms of the local coordinates in the Minkowski spacetime,
in terms of global coordinates.
This can be accomplished by using the inverse vierbeins
$e_{\bar{a}}^{\mu}$ through the relation $\gamma^{\mu}=e_{\bar{a}}^{\mu}\gamma^{\bar{a}}$  $(\mu,\bar{a}=0,1,2,3)$,
with $\gamma^{\bar{a}}=\left(\gamma^{\bar{0}},\gamma^{\bar{i}}\right)$
being the standard gamma matrices.
The equation of motion
governing the dynamics of this system is the modified Dirac--Pauli equation
in the curved space
\begin{equation}
\left[ i\hbar \gamma^{\mu}\left( \partial_{\mu}+\Gamma_{\mu}\right) -
\frac{\mu}{2c}\sigma^{\mu \nu}F_{\mu \nu}-Mc\right] \Psi =0,
\label{diracsc}
\end{equation}
with $\sigma^{\mu \nu}={}i[\gamma^{\mu},\gamma^{\nu}]/2$,
$\left(F_{0i},F_{ij}\right) ={}\left( E^{i},\epsilon_{ijk}B^{k}\right)$,
$\left(\sigma^{0j},\sigma^{ij}\right) ={}\left( i\alpha^{j},
  -\epsilon_{ijk}\Sigma^{k}\right) $,
where $E^{i}$ and $B^{k}$ are the electric and
magnetic field strengths and $\Sigma^{k}$ is the spin operator.
Here, we use the same vierbein of the Ref. \cite{PRD.2008.78.064012},
where the spinorial affine connection $\Gamma_{\mu}$ has been calculated
in detail.
Moreover, in this work, we are only interested on the planar dynamics of a
spin--1/2 neutral particle under the action of a radial electric field.
In this manner we require that $p_{z}=z=0$ and $B^{k}=0$
for $k=1,2,3$.
Furthermore, according to the tetrad postulated \cite{Book.2012.Lawrie},
the matrices $\gamma^{\bar{a}}$ can be any set of constant Dirac
matrices in a such way that we are free to choose a representation for
them.
We choose to work in a representation in which the Dirac matrices are
given in terms of the Pauli matrices, namely
\cite{NPB.1988.307.909,NPB.1989.328.140}
\begin{equation}
  \beta = \gamma^{\bar{0}} = \sigma^{3},\qquad
  \gamma^{\bar{1}} =  i\sigma^{2},\qquad
  \gamma^{\bar{2}} = -is\sigma^{1},
  \label{gmp}
\end{equation}
where $(\sigma^{1},\sigma^{2},\sigma^{3})$ are the Pauli matrices and $s$
is twice the spin value, with $s=+1$ for spin ``up'' and $s=-1$ for spin
``down''.
In this representation, the only non-vanishing
component of the spinorial affine connection $\Gamma_{\mu}$ is found to be
\begin{equation}
  \Gamma_{\varphi}=-i\frac{\left( 1-\alpha \right)}{2}s\sigma^{z}.
\label{newconx}
\end{equation}
For the field configuration, we consider the electric field due to a
linear charge distribution, superposed to the cosmic string.
The expression for this field is seem to be
\begin{equation}
  E_{r}=\frac{2 \lambda}{\alpha r}.
\end{equation}
Therefore, the second order equation associated with Eq. \eqref{diracsc}
reads
\begin{equation}
\hat{H} \Phi = k^2\Phi,
\end{equation}
with
\begin{align}
  \hat{H} = {}
  &
    -\nabla_{\alpha}^{2}
    -\frac{(1-\alpha)s\sigma^{z}}{i\alpha^{2}r^{2}}\partial_{\varphi}
    +\frac{(1-\alpha)^{2}}{4\alpha^{2}r^{2}}
    +\frac{2 \mu s}{\hbar c}\frac{E_{r}}{i \alpha r}\partial_{\varphi}
     \nonumber \\
  &
  +\frac{\mu}{\hbar c} (\partial_{r} E_{r})\sigma^{z}
  -\frac{\mu}{\hbar c}\frac{(1-\alpha)E_{r}}{\alpha r}\sigma^{z}
  +\frac{\mu^{2}}{\hbar^2c^2}E_{r}^{2},
\label{eq:diracso}
\end{align}
where
$\nabla_{\alpha}^{2}=\partial_{r}^{2}+(1/r) \partial_{r}
+(1/\alpha^2 r^2)\partial_{\varphi}^{2}$
is the Laplace-Beltrami operator in the conical space and
$k^{2}=\left(\mathcal{E}^2-M^2c^4\right)/\hbar^{2}c^2$.
As the angular momentum $\hat{J}=-i\partial_{\varphi}+(s/2)\sigma^{z}$,
commutes with the $\hat{H}$, it is possible to decompose the
fermion field as
\begin{equation}
  \Phi=
  \begin{pmatrix}
    \psi \\
    \chi
  \end{pmatrix}=
  \begin{pmatrix}
    \sum_{m} f_{m}(r)\,e^{im\varphi} \\
    \sum_{m} g_{m}(r)\,e^{i\left( m+s\right) \varphi}
    \label{eq:ansatz}
  \end{pmatrix},
\end{equation}
where $m=0,\pm 1,\pm 2,\pm 3,\ldots $ is the angular momentum quantum
number.
In this manner, the radial equation for $f_{m}(r)$ is
\begin{equation}
  hf_{m}(r) = k^{2} f_{m}(r) ,  \label{eq:eigen}
\end{equation}
with
\begin{equation}
  h=h_{0}+\frac{\eta}{\alpha}\frac{\delta (r)}{r},
  \label{eq:hfull}
\end{equation}
and
\begin{equation}
  h_{0}=-\frac{d^{2}}{dr^{2}}-\frac{1}{r}\frac{d}{dr}
  +\frac{j^{2}}{r^{2}},  \label{eq:hzero}
\end{equation}
where
\begin{equation}
  j=\frac{m+s\eta}{\alpha}-\frac{s(1-\alpha)}{2\alpha},
\end{equation}
is the effective angular momentum and
\begin{equation}
  \eta =\frac{\phi}{\phi_{0}}.
\end{equation}
Here, $\phi=4\pi\lambda$ is the electric flux of the electric field and
$\phi_{0}=hc/\mu$ is the quantum of electric flux.

As far as we know, only the scattering problem for the Hamiltonian in
Eq. \eqref{eq:hfull} has been studied in Ref. \cite{JHEP.2004.2004.16}.
However, there, the spin effect was not taken into account once the
author imposed the regularity of the wave function at the origin.
The inclusion of spin gives rise to the Dirac delta function potential,
which comes from the interaction between the spin and the line of
charge, and its inclusion has effects on the scattering phase shift,
giving rise to an additional scattering phase shift
\cite{PRL.1990.64.503}. 
Thus, the main aim of this work is to show that there are bound states
due to the presence of the Dirac delta function.
The approach adopted here is that of the self--adjoint extensions
\cite{Book.2004.Albeverio}, which  has been used to deal with singular
Hamiltonians, for instance, in the study of spin 1/2 Aharonov-Bohm
system and cosmic strings \cite{EPJC.2014.74.2708,PRD.1989.40.1346}, in
the Aharonov-Bohm-Coulomb problem
\cite{EPJC.2013.73.1,TMP.2009.161.1503,PRD.1994.50.7715,JMP.1995.36.5453},
and in the equivalence between the self--adjoint extension and
normalization \cite{Book.1995.Jackiw}.

\section{Scattering and bound states analysis}
\label{bound-scatt}

In this section, we obtain the S-matrix and from its poles an expression
for the bound states is obtained.
Before we solve Eq. \eqref{eq:eigen}, let us first analyze the
Hamiltonian $h_{0}$.

In the von Neumann theory of self--adjoint extensions, a Hermitian
operator $\hat{O}$ ($\hat{O}=\hat{O}^{\dagger}$) defined in a dense
subset of a Hilbert space has deficiency indices $(n_{+},n_{-})$, which
are the sizes of the deficiency sub-spaces spanned by the solutions for
\begin{equation}
  \hat{O} \chi_{\pm} = \pm i \chi_{\pm}.
\end{equation}
When the dimension of the deficiency subspace are zero, the operator is
self--adjoint and it has no additional self--adjoint extension.
When the dimension of the deficiency spaces are not zero the operator is
not self--adjoint.
If $n_{+}=n_{-}=n$ the operator admits a self--adjoint extension
parametrized by a $n \times n$ unitary matrix.
However, if the deficiency indices are not equal, the operator has no
self--adjoint extensions.
By standard results, it is well-known that the Hamiltonian $h_{0}$ has
deficiency indices $(1,1)$ and it is self--adjoint for $|j| \geq 1$,
whereas for $|j| < 1$ it is not self--adjoint, and admits an
one-parameter family of self--adjoint extensions
\cite{Book.1975.Reed.II}.
Actually, $h$ can be interpreted as a self--adjoint extension of $h_{0}$
\cite{crll.1987.380.87}.
All the self--adjoint extension of $h_{0}$, $h_{0,\nu}$, are accomplished
by requiring the boundary condition at the origin \cite{JMP.1985.26.2520}
\begin{equation}
  \nu f_{0,j} = f_{1,j},
  \label{eq:bc}
\end{equation}
where $-\infty < \nu \leq \infty$ and $-1 < j < 1$.
The boundary values are
\begin{align*}
  f_{0,j} = {}
  & \lim_{r \to 0^{+}}r^{|j|}f_{m}(r),  \nonumber \\
  f_{1,j} = {}
  & \lim_{r \to 0^{+}}\frac{1}{r^{|j|}}
    \left[
    f_{m}(r)-f_{0,j}\frac{1}{r^{|j|}}\right].
\end{align*}
In Eq. \eqref{eq:bc} $\nu$ is the self--adjoint extension parameter.
It turns out that $1/\nu$ represents the scattering
length of $h_{0,\nu}$ \cite{Book.2004.Albeverio}.
For $\nu=\infty$ (the Friedrichs extension of $h_{0}$), one has the free
Hamiltonian (without spin) with regular wave functions at the origin
($f_{m}(0)=0$).
This situation is equivalent to impose the Dirichlet boundary condition
on the wave function.
On the other hand, if $|\nu|<\infty$, $h_{0,\nu}$ describes a point
interaction at the origin.
In this latter case the boundary condition permits
a $r^{-|j|}$ singularity in the wave functions at the origin
\cite{PRA.2008.77.036101}.

\begin{figure}
  \centering
  \includegraphics[width=\columnwidth]{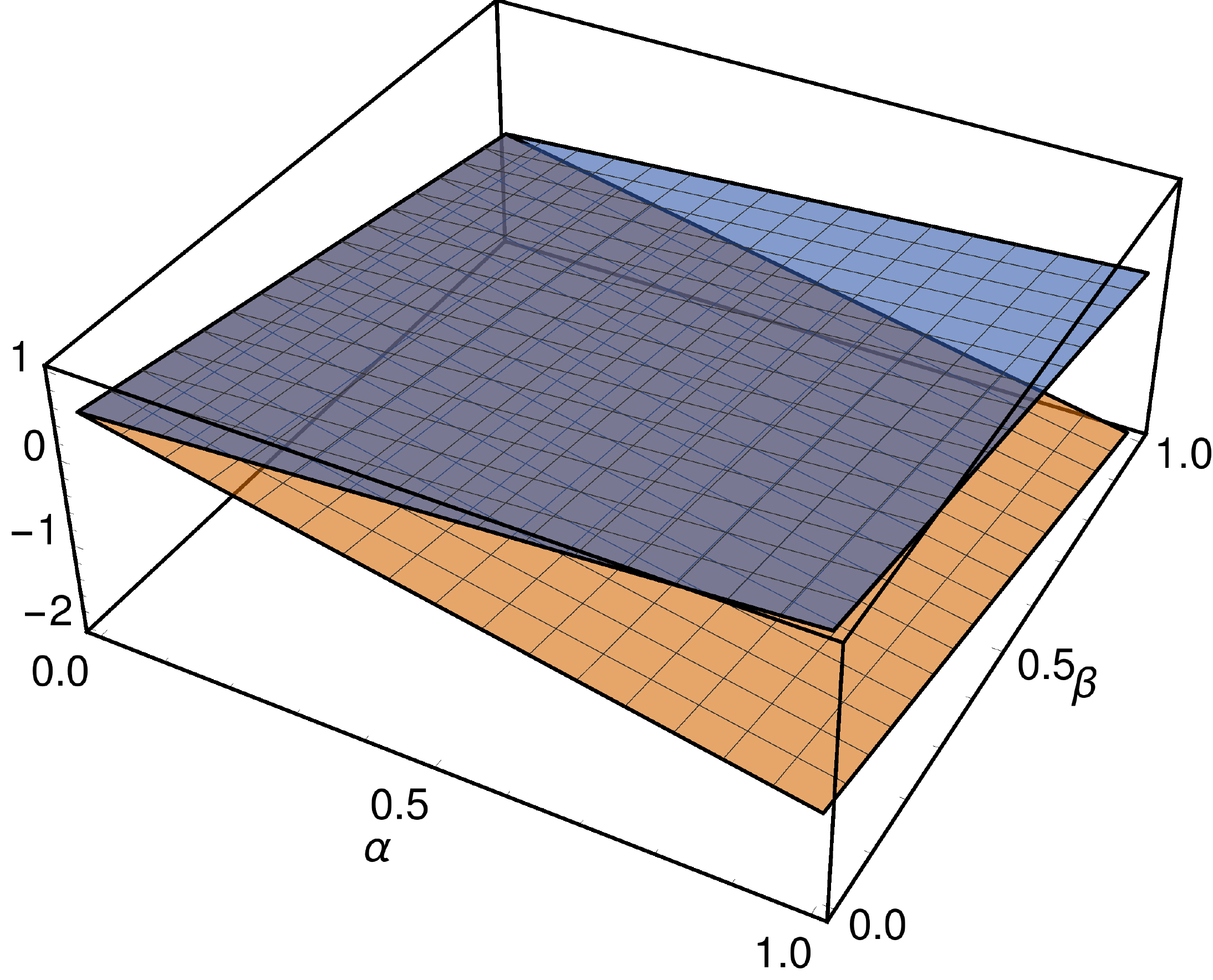}
  \caption{In this figure we plot the planes
    $\pi_{-}(\alpha,\beta)$ (orange, bottom) and
    $\pi_{+}(\alpha,\beta)$ (blue, top)
    for $N=0$ and $s=+1$.
    The region between the two planes is that in which the operator $h_0$
    is not self--adjoint.
   }
  \label{fig:fig1}
\end{figure}

\begin{figure*}
  \centering
  \includegraphics[width=\columnwidth]{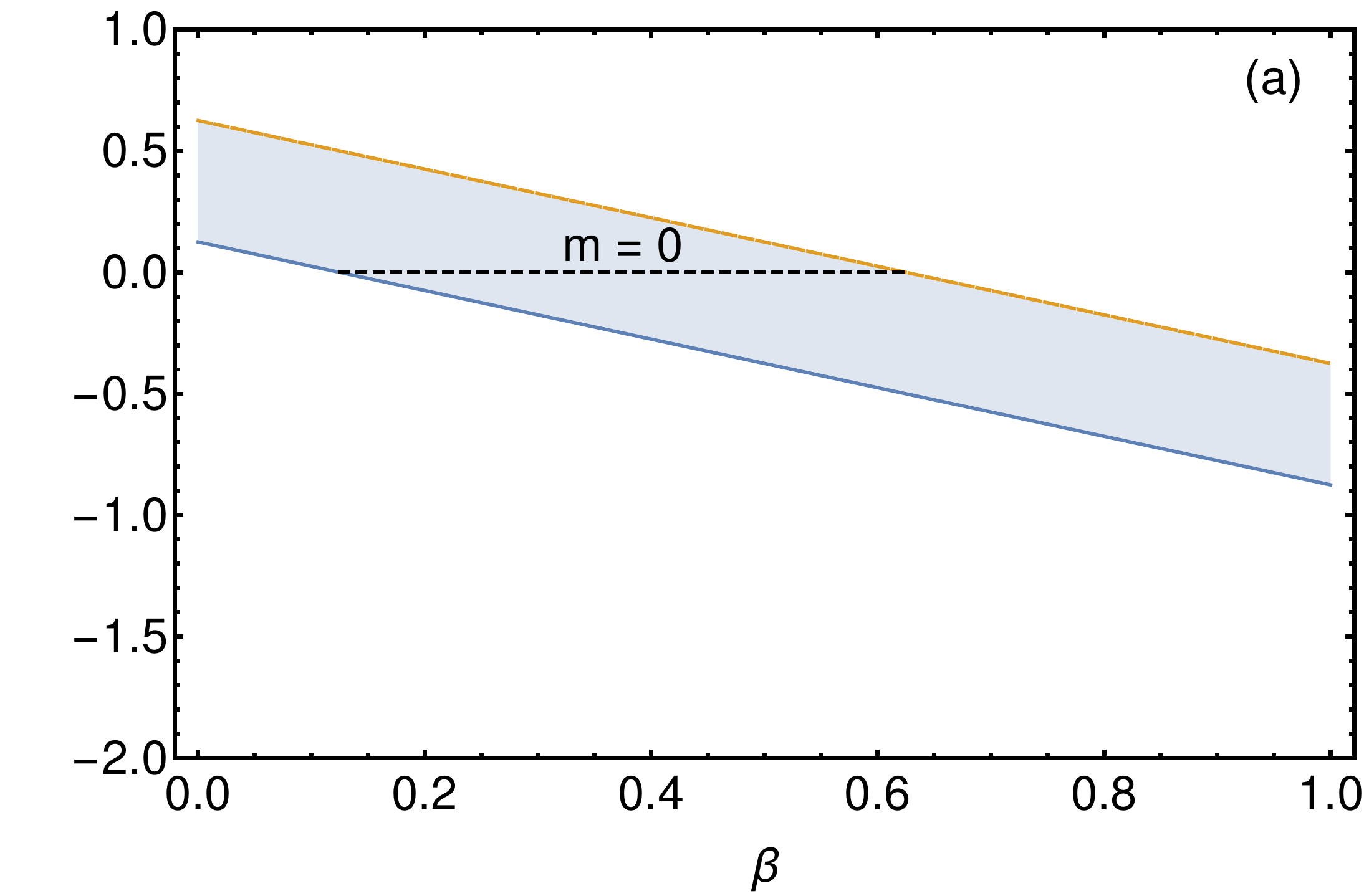}%\hspace{0.5cm}
  \includegraphics[width=\columnwidth]{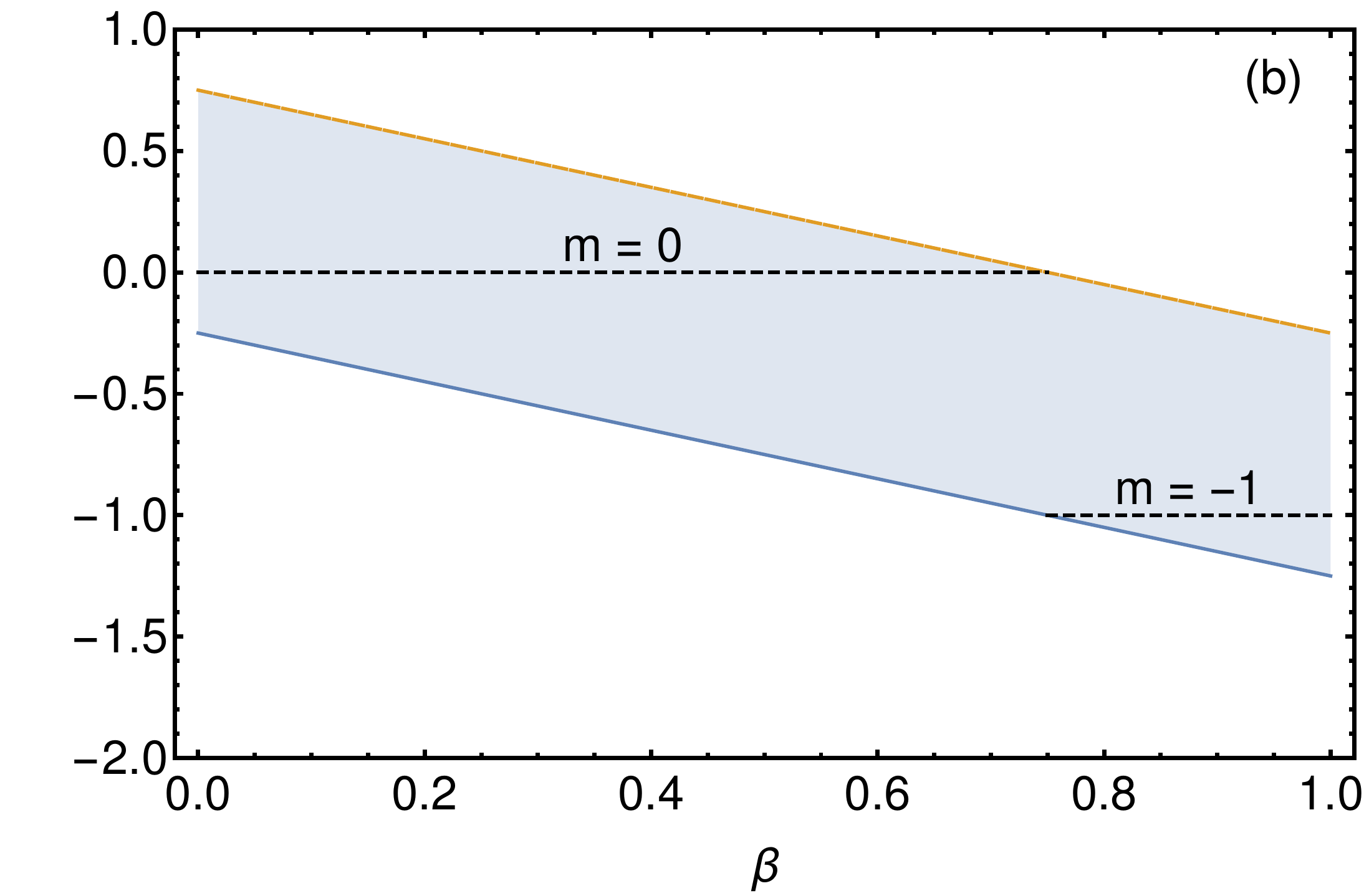}\\
  \vspace{0.5cm}
  \includegraphics[width=\columnwidth]{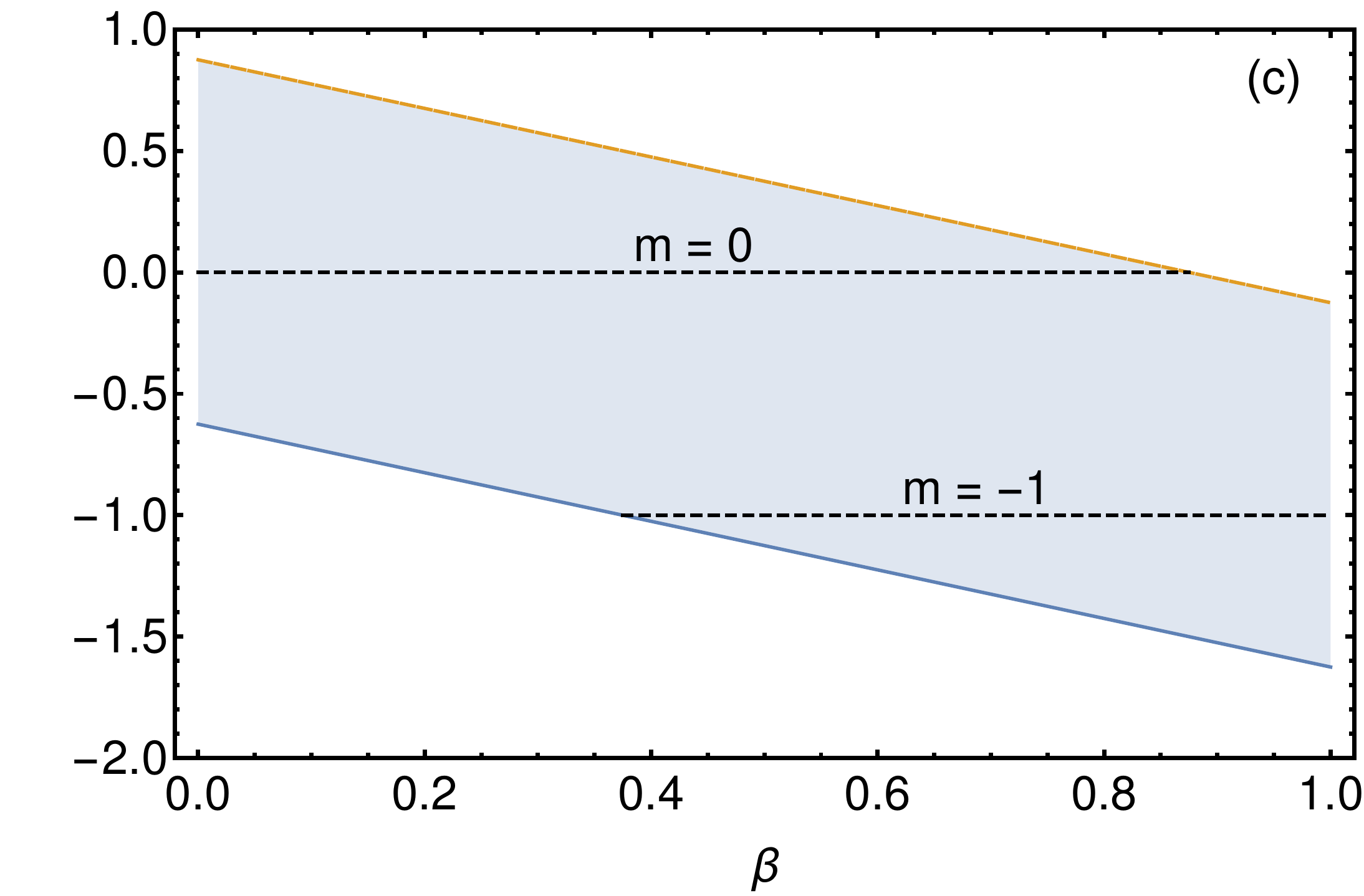}%\hspace{0.5cm}
  \includegraphics[width=\columnwidth]{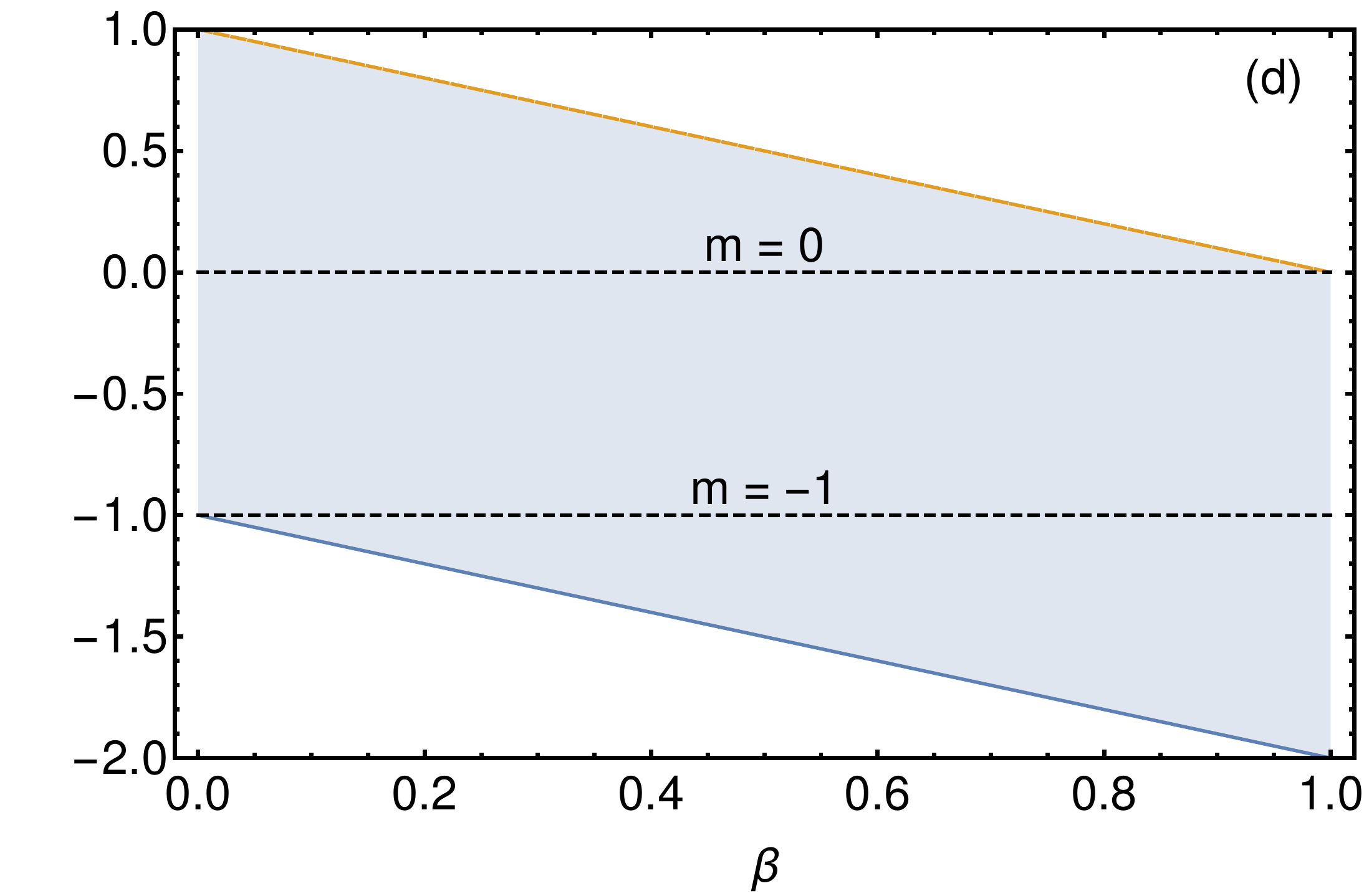}
  \caption{Cross sections of Fig. \ref{fig:fig1} for
    different values of the deficit angle: (a) $\alpha=0.25$, (b)
    $\alpha=0.50$, (c) $\alpha = 0.75$ and (d) $\alpha=1.00$.
    The shaded area schematically represents that area in which the
    operator $h_0$ is not self--adjoint.
    The dashed lines represent the values of angular moment
    quantum number.
 }
  \label{fig:fig2}
\end{figure*}

Let us now discuss for which values of the angular moment quantum number
$m$, the operator $h_{0}$ is not self--adjoint.
In fact, these values depending on the variables $\alpha$ and $\eta$.
As discussed in \cite{JHEP.2004.2004.16}, $0 <\alpha<1$ represents a
positive curvature and a planar deficit angle, corresponding to a
conical spacetime.
On the other hand, $\alpha>1$ represents a negative curvature and
an excess of planar angle, corresponding to an anti-conical spacetime.
Finally, $\alpha=1$ corresponds to a flat space.
Then, we focus on a conical spacetime.
For the electric flux $\eta$ let us adopt the decomposition defined by
\cite{PRD.41.2015.1990}
\begin{equation}
  \eta = N + \beta,
\end{equation}
being $N$ an integer and
\begin{equation}
  0 \leq \beta < 1.
\end{equation}
The inequality $|j|<1$ then reads
\begin{equation}
  \pi_{-}(\alpha,\beta)
  < m <
  \pi_{+}(\alpha,\beta),
\end{equation}
with
$\pi_{\pm}(\alpha,\beta)=\pm\alpha - [(2 \beta+\alpha-1)s/2-s N]$.
In Fig. \ref{fig:fig1} we plot the planes
$\pi_{\pm}(\alpha,\beta)$ for $N=0$ and $s=+1$.
The region between these two planes is that in which the operator
$h_{0}$ is not self--adjoint.
In Fig. \ref{fig:fig2} we show cross sections of this region for
some particular values of the deficit angle $\alpha$.
We can observe in Fig. \ref{fig:fig2}(a) that, for $\alpha=0.25$, only
for $m=0$ the operator $h_0$ is not self adjoint, whereas for
$\alpha=0.50$ (Fig. \ref{fig:fig2}(b)) the
operator $h_{0}$ is not self adjoint for $m=0$ and $m=-1$, but not for
both values of $m$ at same time for the whole range of $\beta$ values.
Indeed, a necessary condition for the operator $h_{0}$ not being
self--adjoint for the state with $m=-1$ is $\alpha > 1 - 2 \beta/3$.
In fact, this condition is also valid for $N \neq 0$ and, in this latter
case, the $m$ values for which the $h_{0}$ is not self--adjoint are
shifted to the values $s N$ and $s N - 1$.
For $\alpha=0.75$ (Fig. \ref{fig:fig2}(c)) we can observe that there is
a range the values of $\beta$ in which, for both values of $m=0$ and
$m=-1$, the operator $h_0$ is not self--adjoint.
And last but not least, $\alpha=1.0$ (see Fig \ref{fig:fig2} (d)) is the
only situation in which the operator $h_0$ is not self--adjoint for the
both values of angular momentum quantum number for the whole range of
$\beta$ (the unique exception is $\beta=0$).

Now, let us comeback to the solution of Eq. \eqref{eq:eigen}.
As a matter of fact, it is the Bessel differential equation.
Thus, the general solution for $r\neq 0$ is seen to be
\begin{equation}
  f_{m}(r)=a_{m}J_{|j|}(kr)+b_{m}J_{-|j|}(kr),  \label{eq:Bessel}
\end{equation}
where $J_{\nu}(z)$ is the Bessel function of fractional order.
The coefficients $a_{m}$ and $b_{m}$ represent the contributions of the
regular and irregular solutions at the origin, respectively.
Thus making use of the boundary condition in Eq. \eqref{eq:bc} in the
subspace $|j| < 1$, a relation between the coefficients is
obtained, namely,
\begin{equation}
  b_{m}=-\mu_{\nu} a_{m},
    \label{eq:relation}
\end{equation}
where the term $\mu_{\nu}$ is given by
\begin{equation}
  \mu_{\nu} =
   \frac
  {k^{2|j|}\Gamma(1-|j|)\sin(|j| \pi)}
  {4^{|j|}\Gamma(1+|j|) \nu+ k^{2|j|}\Gamma(1-|j|)\cos(|j|\pi)},
  \label{eq:mu}
\end{equation}
where $\Gamma(z)$ is the gamma function.
Therefore, in this subspace the solution reads
\begin{equation}
  f_{m}(r)=a_{m}\left[J_{|j|}(kr)-\mu_{\nu} J_{-|j|}(kr)\right].
\end{equation}
The above equation shows that the self--adjoint extension parameter $\nu$
controls the contribution of the irregular solution $J_{-|j|}$ for the
wave function.
As a result, for $\nu=\infty$, we have $\mu_{\infty}=0$, and there is no
contribution of the irregular solution at the origin for the wave
function.
Consequently, the total wave function reads
\begin{equation}
  \psi = \sum_{m=-\infty}^{\infty} a_m J_{|j|}(k r) e^{i m \varphi}.
  \label{eq:regpsi}
\end{equation}
It is well-known that the coefficient $a_{m}$ must be chosen in a such way
that $\psi$ represents a plane wave that is incident from the right.
In this manner, we obtain the result
\begin{equation}
  a_{m} = e^{-i|j| \pi/2}.
\end{equation}
The scattering phase shift can be obtained from the asymptotic behavior
of Eq. \eqref{eq:regpsi}.
This leads to
\begin{equation}
  \delta_{m}= \frac{\pi}{2}(|m|-|j|).
  \label{eq:regphaseshift}
\end{equation}
This is the scattering phase shift of the Aharonov-Casher effect in the
cosmic string background.
It is worthwhile to note that, for $\alpha=1$, it reduces to the phase
shift for the usual Aharonov-Casher effect in flat space
$\delta_{m}=\pi(|m|-|m+s\eta|)/2$ \cite{PRL.1984.53.319}.

On the other hand, for $|\nu| < \infty$, the contribution of the
irregular solution modifies the scattering phase shift to
\begin{equation}
  \delta_{m}^{\nu}=
  \delta_{m}+\arctan (\mu_{\nu}).  \label{eq:phaseshift}
\end{equation}
Thus one obtains
\begin{align}
  S_{m}^{\nu}
  =
  e^{2 i \delta_{m}^{\nu}}
  =
  e^{2 i \delta_{m}}
  \left(
  \frac{1+i\mu_{\nu}}{1-i\mu_{\nu}}
  \right),
  \label{eq:Sm}
\end{align}
which is the expression for the S-matrix in terms of the phase shift.
As a result, one observes that in this latter case there is an
additional scattering for any value of the self--adjoint extension
parameter $\nu$.
When $\nu=\infty$, we have the S-matrix for the Aharonov-Casher effect on
the cosmic string background, as it should be.

The S-matrix or scattering matrix relates incoming and outgoing wave
functions of a physical system undergoing a scattering process.
Bound states are identified as the poles of the S-matrix in the upper
half in the complex $k$ plane.
In this manner, the poles are determined at the zeros of the denominator
in Eq. \eqref{eq:Sm} with the replacement $k\rightarrow i\kappa $ with
$\kappa=\sqrt{-\left(\mathcal{E}^{2}-M^{2}c^{4}\right)/\hbar^{2}c^{2}}$.
Therefore, for $\nu<0$, one can determine that the present system has
a bound state with energy given
\begin{equation}
  \mathcal{E}=\pm \sqrt{M^{2}c^{4}-4\hbar^{2}c^{2}
  \left[-\nu \frac{\Gamma(1+|j|)}{\Gamma(1-|j|)}\right]^{1/|j|}}.
\label{eq:energy_BG-sc}
\end{equation}
and the normalized radial bound state wave function is
\begin{align}
  \begin{pmatrix}
    f_{m}(r)\\
    g_{m}(r)
  \end{pmatrix} = {}
& 
  \sqrt{\frac{2 \alpha \kappa^{2}/\pi}
  {|j| \alpha \csc(|j|\pi)+|j'| \csc(|j'|\pi)}}
  \nonumber \\                
& \times
  \begin{pmatrix}
    K_{|j|}(\kappa r)\\
    K_{|j'|}(\kappa r)
  \end{pmatrix},
    \label{eq:wavefunction}
\end{align}
where $j'= j+(s/\alpha)(2-\alpha)$ and $K_{\nu}(z)$ is the modified Bessel function of the second kind.
So, there are bound states when the self--adjoint extension parameter is
negative.
In the non-relativistic limit and for $\alpha=1$,
Eq. \eqref{eq:energy_BG-sc} coincides with the bound state energy found in
Ref. \cite{EPJC.2013.73.2402} for the Aharonov-Casher effect in the flat
space.

As a result, it is possible to express the S-matrix in terms of the bound
state energy.
The result is seem to be
\begin{equation}
  S_{m}^{\nu} = e^{2i\delta_{m}}
  \left[
    \frac
    {e^{ 2 i \pi |j|}-(\kappa/k)^{2 |j|}}
    {1-(\kappa/k)^{2 |j|}}
  \right].
\end{equation}

Once we have obtained the S-matrix, it is possible to write down the
scattering amplitude $f(k,\varphi)$.
The result is
\begin{align}
  f(k,\varphi)
  = {}
  &
    \frac{1}{\sqrt{2\pi ik}}
    \sum_{m=-\infty}^{\infty}\left(S_{m}^{\nu}-1\right)
    e^{i m \varphi} \nonumber \\
  = {}
  &
    \frac{1}{\sqrt{2\pi i k}}
    \Bigg\{
    \sum_{m\in\{|j| \geq 1\}}\left(e^{2i\delta_{m}}-1\right)
    e^{i m  \varphi}\nonumber \\
  & +
    \sum_{m\in\{|j| <1\}}
    \left[e^{2i\delta_{m}}
    \left(
    \frac{1+i\mu_{\nu}}{1-i\mu_{\nu}}
    \right)
    -1\right] e^{i m \varphi}
    \Bigg\}.
  \label{eq:scattamp}
\end{align}
In scattering problems the length scale is set by $1/k$, thus the
scattering amplitude $f(k,\varphi)$ would be a function of angle alone,
multiplied by $1/k$ \cite{PRD.1977.16.1815}.
However, we observe that $f(k,\varphi)$ has a dependence
on $\mu_{\nu}$, which in its turn has explicit dependence on $k$
(see Eq. \eqref{eq:mu}).
This behavior is associated with the failure of helicity conservation.
The helicity operator, defined by
\begin{equation}
  \hat{h}=\boldsymbol{\Sigma} \cdot (-i\boldsymbol{\nabla}_{\alpha}
  -e \mathbf{A}),
  \label{eq:helicity}
\end{equation}
obeys the equation
\begin{equation}
 \frac{d\hat{h}}{dt}= e \boldsymbol{\Sigma} \cdot \mathbf{E},
\end{equation}
whit $\boldsymbol{\Sigma}$ is the spin operator and in
Eq. \eqref{eq:helicity}  $\mathbf{A}$ is the potential vector, which
is absent in the present problem.
Therefore, due to the presence of electric field the helicity is not
conserved.

\section{Conclusions}
\label{conclusion}

In this work, we reexamined the relativistic quantum dynamics of a
spin--1/2 neutral particle in the cosmic string spacetime.
This problem has been studied in Ref. \cite{JHEP.2004.2004.16} in the
non-relativistic scenario. However only the scattering solutions were
studied and without taking into account the possibility of bound states.
Here, we have showed that the inclusion of electron spin, which gives
rise to a point interaction, changes the scattering phase shift and
consequently the S-matrix.
The results were obtained by imposing the boundary condition in Eq.
\eqref{eq:bc}, which comes from the von Neumann theory of the
self--adjoint extensions.
Our results are dependent on the self--adjoint extension parameter
$\nu$.
For the special value of $\nu=\infty$ we recover the results of
Ref. \cite{JHEP.2004.2004.16}.
Our expression for the scattering amplitude has an energy dependency.
So, the helicity is not conserved in the scattering process.
Last but not least, examining the poles of the S-matrix, an expression
for the bound state energy was determined.
The presence of bound states has not been discussed before.

\section*{Conflict of interest}
The authors declare that there is no conflict of interests regarding the
publication of this paper.

\section*{Acknowledgements}
FMA thanks Simone Severini and Sougato Bose by their hospitality
at University College London.
This work was partially supported by the CNPq, Brazil,
Grants Nos. 482015/2013--6 (Universal), 476267/2013--7 (Universal),
460404/2014-8 (Universal), 306068/2013--3 (PQ), 311699/2014-6 (PQ), 
FAPEMA, Brazil, Grants No. 01852/14 (PRONEM) and FAPEMIG.

%\section*{References}
\bibliographystyle{apsrev4-1}
%\bibliography{n-string-ahep}

\end{document}